\documentclass[runningheads]{llncs}
\usepackage[T1]{fontenc}
\usepackage{graphicx}
\usepackage{epsfig}
\usepackage{graphicx}
\usepackage{amsmath}
\usepackage{amssymb}
\usepackage{algorithm}
\usepackage{color}
\usepackage{multirow}
\usepackage{booktabs}
\usepackage{array}
\usepackage{setspace}
\usepackage{cite}
\usepackage{tikz,xcolor}
\usepackage{subcaption}
\usepackage{wrapfig}
\usepackage{caption}
\usepackage[backref]{hyperref}
\usepackage{pifont}
\hypersetup{hidelinks}
\usepackage{textcomp}
\usepackage{algpseudocode}
\usepackage{xspace}

\usepackage{marvosym}
\usepackage[switch]{lineno}
\usepackage[marginal]{footmisc}
\usepackage{colortbl}
\usepackage[utf8]{inputenc}
\usepackage{marvosym}
\usepackage{hyperref}
\hypersetup{colorlinks=true,linkcolor=blue,urlcolor=blue,citecolor=blue}
\definecolor{mygray}{gray}{.85}

\begin{document}
\title{CriDiff: Criss-cross Injection Diffusion Framework via Generative Pre-train for Prostate Segmentation}
\titlerunning{Criss-cross Injection Diffusion}

%
% Leiye Liu, Jialong Zhong, Shuyao Wang, Yongri Piao (通信), Huchuan Lu
% \author{
%   Tingwei Liu\inst{1} 
%   \index{Liu, Tingwei}
%   \and
%   Miao Zhang \inst{1}
%   \index{Zhang, Miao} 
%   \and 
%   Leiye Liu \inst{1}
%   \index{Liu, Leiye} 
%   \and
%   Jialong Zhong \inst{1}
%   \index{Zhong, Jialong} 
%   \and 
%   Shuyao Wang \inst{1} 
%   \index{Wang, Shuyao} 
%   \and 
%   Yongri Piao \inst{1}\textsuperscript{(\Letter)}
%   \index{Piao, Yongri} 
%   \and 
%   Huchuan Lu \inst{1}
%   \index{Lu, Huchuan} 
% }
\author{
  Tingwei Liu
  \index{Liu, Tingwei}
  \and
  Miao Zhang
  \index{Zhang, Miao} 
  \and 
  Leiye Liu
  \index{Liu, Leiye} 
  \and
  Jialong Zhong
  \index{Zhong, Jialong} 
  \and 
  Shuyao Wang
  \index{Wang, Shuyao} 
  \and \\ 
  Yongri Piao \textsuperscript{(\Letter)}
  \index{Piao, Yongri} 
  \and 
  Huchuan Lu
  \index{Lu, Huchuan} 
}
\authorrunning{T. Liu et al.}
\institute{Dalian University of Technology, China\\
% tingweiliu@mail.dlut.edu.cn, \{zhanglihe,lhchuan\}@dlut.edu.cn
% }
% \email{\{tingweiliu,flameliu,jialongzhong,dlutwsy\}@mail.dlut.edu.cn, \{miaozhang,yrpiao,lhchuan\}@dlut.edu.cn}}
\email{tingweiliu@mail.dlut.edu.cn, yrpiao@dlut.edu.cn}}

\maketitle              % typeset the header of the contribution
\begin{abstract}
Recently, the Diffusion Probabilistic Model (DPM)-based methods have achieved substantial success in the field of medical image segmentation.
However, most of these methods fail to enable the diffusion model to learn edge features and non-edge features effectively and to inject them efficiently into the diffusion backbone.
Additionally, the domain gap between the images features and the diffusion model features poses a great challenge to prostate segmentation.
In this paper, we proposed CriDiff, a two-stage feature injecting framework with a Crisscross Injection Strategy (CIS) and a Generative Pre-train (GP) approach for prostate segmentation. 
The CIS maximizes the use of multi-level features by efficiently harnessing the complementarity of high and low-level features. 
To effectively learn multi-level of edge features and non-edge features, we proposed two parallel conditioners in the CIS: the Boundary Enhance Conditioner (BEC) and the Core Enhance Conditioner (CEC), which discriminatively model the image edge regions and non-edge regions, respectively.
Moreover, the GP approach eases the inconsistency between the images features and the diffusion model without adding additional parameters.
Extensive experiments on four benchmark datasets demonstrate the effectiveness of the proposed method and achieve state-of-the-art performance on four evaluation metrics. 
The source code will be publicly available at \href{https://github.com/LiuTingWed/CriDiff}{https://github.com/LiuTingWed/CriDiff}.
\end{abstract}
\section{Introduction}
Prostate cancer, as the second most common cancer affecting men, necessitates accurate diagnostic tools for effective management\cite{siegel2020cancer}. 
Precise segmentation of the prostate is critical for the diagnosis and treatment planning of prostate cancer.
% Precise delineation of the prostate from magnetic resonance imaging (MRI) and transrectal ultrasound (TRUS) is crucial for the diagnosis and treatment planning of prostate-related diseases, including prostate cancer, prostatitis, benign prostatic hyperplasia, and other prevalent conditions affecting the prostate. 
With the development of deep learning, convolutional neural networks (CNNs) have made significant progress for prostate segmentation\cite{pellicer2022deep, tian2015superpixel, guo2015deformable}. 
% Among them,  Pellicer-Valero \textit{et al.}\cite{pellicer2022deep} proposed a Retina U-Net detection framework to locate and segment the prostate lesions.
% Tian \textit{et al.}\cite{tian2015superpixel} proposed a superpixel-based 3D graph cut algorithm by combining a 3D  cuts and a 3D active contour model for segmenting the prostate MR images. 
% Guo \textit{et al.}\cite{guo2015deformable} introduced a deformable method for prostate segmentation, applying sparse patch matching to generate a prostate likelihood map.
Although the above methods achieve promising results, they use the softmax in the cross-entropy loss overemphasizes the highest logit, leading to deterministic predictions.
% they adpoted the powerful softmax in the cross-entropy loss, which highly promotes the probability of the highest logit, predicting deterministic results.
% even on wrong predictions
However, estimating the model's output uncertainty is crucial for clinical doctors to further diagnose uncertain areas, because medical segmentation problems are often characterized by ambiguities and multiple hypotheses may be plausible\cite{warfield2002validation}.
\par
Recently, Diffusion Probabilistic Models (DPMs) have led to unprecedented advancements in content generation tasks. 
Because they have the capability to generate different predictions by running multiple times, many DPM-based methods\cite{wolleb2022ensemble, bozorgpour2023dermosegdiff, wu2023medsegdiff2, wu2023medsegdiff} are proposed in the field of medical image segmentation.
% Inspired by its success, many DPM-based methods\cite{wolleb2022ensemble, bozorgpour2023dermosegdiff, wu2023medsegdiff2, wu2023medsegdiff} are proposed in the field of medical image segmentation, achieving good performance, because they have the capability to generate different segmentation predictions by running multiple times. 
% Wolleb \textit{et al.}\cite{wolleb2022ensemble} proposed EnsDiff that adopts input images as priors to generate segmentation distributions, creating uncertainty maps and an implicit ensemble of segmentations. 
% Wu \textit{et al.}\cite{wu2023medsegdiff} proposed MedSegDiffV1 with dynamic conditional encoding and FF-Parser module to mitigate high-frequency noise effects. 
% MedSegDiffV2 \cite{wu2023medsegdiff2} enhances V1 with SS-Former module to model the segmentation noise and semantic feature interaction.
% Despite these methods have shown great performances in the medical image segmentation, the intricate anatomical positioning of the prostate, along with its visual similarity to adjacent tissues, presents significant challenges on accurate prediction of edge\cite{yu2017volumetric}.
Despite these methods have shown great performances, the intricate anatomical positioning of the prostate, along with its visual similarity to adjacent tissues, presents significant challenges on accurate prediction of edge\cite{yu2017volumetric}.
However, these methods overlook the learning of boundary information and treat all regions with equal importance.
DermoSegDiff \cite{bozorgpour2023dermosegdiff} introduces a novel boundary loss function by calculating the distance between each foreground pixel in the ground-truth label and the nearest background pixel.
% Despite achieving good performance, this weighted loss function not only depends on the accuracy of label but also places less emphasis on non-edge regions.
However, this weighted loss requires careful adjustment on the coefficients to balance the learning between edge and non-edge areas, relying on laborious trial and error.
Moreover, previous DPM-based methods inject the multi-level features of medical images stage by stage into the diffusion backbone (\textit{e.g.}, high-level semantic features are injected into deeper layers and low-level features are injected to shallower layers).
This approach leads to the underutilization of multi-level features, limiting early-stage accuracy in object localization or shaping and impeding the model's capability to generate fine-grained objects in later stages.
% In other words, the strategy of merely injecting high-level semantic features into deeper layers limits early-stage accuracy in object localization or shape. 
% Conversely, injecting low-level features to shallower layers hinders the model's capability to generate fine-grained objects in later stages.
% However, directly injecting small-scale high-level features into shallow layers or large-scale low-level features into deep layers causes scale mismatches, impairing learning and performance.
% Therefore, it is essential to design a structure that explicitly models boundary information while paying attention to non-edge areas, and to devise a strategy that maximizes the utilization of multi-level features.
% Therefore, it is essential to design a strategy that 
Therefore, it is essential to design a strategy that effectively learns multi-level features of edges and non-edges and enhances utilization of these features when integrating them into the diffusion model.
\par
When the diffusion model applies to segmentation tasks, randomly initialized diffusion model parameters diffuse the final prediction map under the guidance of image conditional features from the specific data domain.
% Furthermore, the original diffusion model is initially proposed for the image generation. When applied to segmentation tasks, randomly initialized diffusion model features diffuse the final prediction map under the guidence of image conditional features from the specific data domain.
The difference between diffusion model features and conditional features creates a domain gap, especially pronounced in prostate images. This domain gap impedes model convergence and diminishes performance.
% The difference of the diffusion model features and the conditional features results in a domain gap. This domain gap particularly pronounced in medical images such as prostate images, impedeing model convergence and diminishes performance.
% The difference of the diffusion model features and  the domain gap between the conditional features and the diffusion model features results in a discordance, particularly pronounced in medical images such as prostate images.
% This discordance impedes model convergence and diminishes performance.
% Stable Diffusion (SD)\cite{rombach2022high}, trained on numerous natural images, provides robust pre-trained parameters for diffusion models, which could potentially narrow the domain gap. 
% Directly applying the SD pre-trained model to the prostate segmentation might lead to over-convergence, owing to discrepancies in data distribution.
Current DPM-based approaches in medical image segmentation have not fully considered this issue.
% do not fully capitalize on the potential of the diffusion model for feature learning.
Consequently, it is essential to introduce an efficient method to reduce the domain gap, enabling better feature learning in diffusion models.
\par
% To this end, we strive to confront challenges of achieving accurate prostate segmentation.
% The first challenge towards this goal is to design a structure capable of decoupling images feature, learning both image edges and non-edge areas with equal importance, and efficiently injecting these features into the diffusion backbone. 
% The second challenge entails proficiently extracting the diffusion model's potential for feature learning through a streamlined and effective approach
To address the aforementioned problems, we proposed a novel two-stage framework with a feature injection strategy and a generative pre-train method for prostate segmentation, entitled CirDiff. 
Specifically, we proposed the Crisscross Injection Strategy (CIS) for enabling the diffusion to complementarily utilize multi-level features of edges and non-edge areas.
To this goal, we proposed two parallel conditioners in this strategy, named Boundary Enhance Conditioner (BEC) and Core Enhance Conditioner (CEC).
These two conditioners with distinct structures, are capable of discriminative learning of image edge features and other non-edge regions features. 
The BEC employs a triangular architecture that progressively cuts down the number of layers, focusing on learning edge textures features. 
Conversely, the CEC focuses on learning non-edge semantic features via an inverted triangular architecture that progressively increases the number of layers. 
Then, we injected them into the diffusion backbone in a crisscross manner, promoting the diffusion model ability of learning edges and non-edge areas.
% To maximize the utilization of these features, the CIS leverages the complementarity of high and low-level features, criss-cross injecting them into the diffusion backbone, thus promoting the accuracy of prostate segmentation.
Furthermore, we introduced a Generative Pre-train (GP) approach for prostate segmentation. 
The GP pretrains the diffusion model on generative tasks within the target domain, aligning feature representations more closely with the target domain. 
This approach narrows the domain gap between the conditional features and the diffusion model features, thus improving model performance without introducing additional parameters.
In brief, the contributions of this paper are: 
% \ding{202} We propoed a novel framework included Boundary Enhance Conditioner (BEC), Core Enhance Conditioner (CEC) and Criss-cross Injection Strategy (CIS) to effective learning the edge and core information of prostate and injecting them to the Diffusion.
(1) We proposed a novel Crisscross Injection Strategy (CIS) with Boundary Enhance Conditioner (BEC) and Core Enhance Conditioner (CEC) to enhance the diffusion model's capability to learning features in both edge and non-edge areas.
(2) We introduced a Generative Pretrain (GP) method for prostate segmentation to reduce the domain gap between the conditional features and the diffusion model features, improving model convergence.
(3) We demonstrated our proposed method achieving SOTA performance on three MRI prostate datasets and one ultrasound prostate dataset under four evaluation metrics.
\section{Methods}
\begin{figure}[t]
  \centering 
  \scalebox{0.4}{\includegraphics{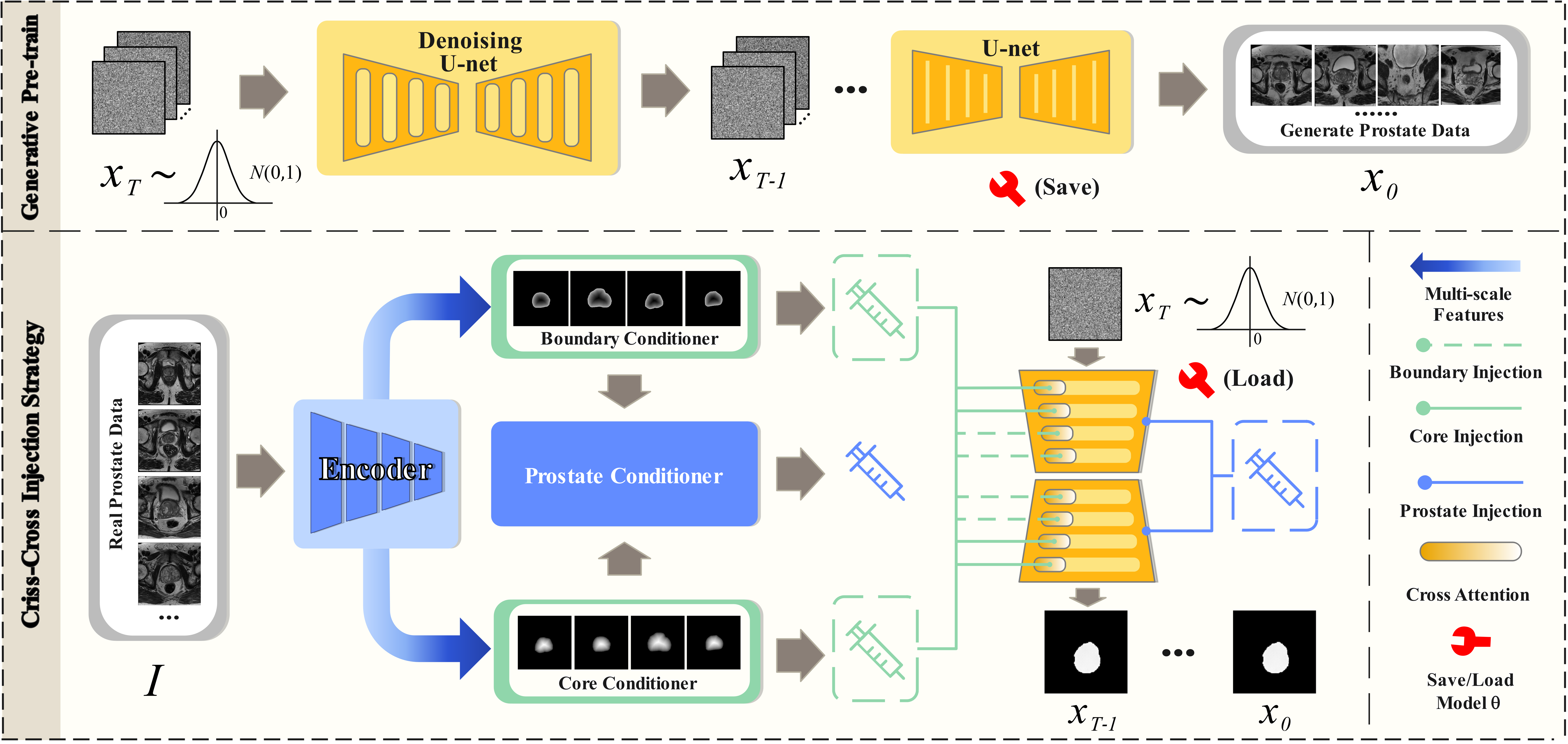}} % Adjust the scaling factor as needed
  \caption{Overall of our method. Up: The first stage is our proposed generative pre-train that is described in Sec. 2.1. Bottom: After pre-training, we performed the criss-cross injection strategy to segment prostate in Sec. 2.2 and Sec. 2.3.} \label{fig1}
\vspace{-0.2cm}
\end{figure}

% The overall architecture of our proposed CriDiff is shown in Figure \ref{fig1}.
The architecture of CriDiff is shown in Figure \ref{fig1}.
In the first stage, we leveraged DPM's generative power to formulate the segmentation task as a generative problem, developing a model that precisely captures the characteristics of prostate images. 
In the second stage, the CIS injects both boundary and core features into the pre-trained diffusion model in a crisscross way.
To effectively learn boundary and core areas features, we employed the proposed BEC and CEC to separately learn the boundary and core features of prostate images, respectively.
Finally, a Gaussian noise is guided by the boundary, core and image feature information to generate the final prediction map.
\subsection{Generative Pre-train Approach}
To reduce the domain gap between the conditional features and the diffusion model features, we introduced a generative pre-train approach for prostate segmentation.
The diffusion model operates through two processes: initially, in the forward process, an image is progressively noised over $T$ steps by adding Gaussian noise. Subsequently, in the reverse stage, a neural network learns to recover the original data by reversing this noise addition.
Given prostate images \( I_{0} \), the reverse process can be represented as: $p_{\theta}(I_{0:T-1} \mid I_T) = \prod_{t=1}^{T} p_{\theta}(I_{t-1} \mid I_t),$
% the parameter \(\theta\) learns the data distribution through a diffusion process. This can be mathematically expressed as:
% \begin{equation}
%   \label{eq1}
%   p_{\theta}(I_{0:T-1} \mid I_T) = \prod_{t=1}^{T} p_{\theta}(I_{t-1} \mid I_t),
% \end{equation}
where $\theta$ represents the denoising model parameters and $p_{\theta}(I_T)$ is the latent variable distribution. 
Through the training process, the parameters $\theta$ acquire the capability to represent prostate features, enabling the transformation from a Gaussian noise distribution to the prostate data distribution $p_{\theta}(I_0)$.
% Some generative prostate images can be found in the supplementary material.
Following \cite{ho2020denoising}, the $\theta$ is considered as a noise prediction network $\epsilon_{\theta}$, optimized by a simple mean-squared error:
\begin{equation}
  \label{eq2}
  \mathcal{L}(\theta) =  \| \epsilon_{\theta}(I_{t}) - \epsilon_{t} \|^2,
  \end{equation}
where $I_{t}$ is a noised prostate image at $t$ step. By applying the reparameterization, $I_t = \sqrt{\hat{\alpha}_t} I_0 + \sqrt{1 - \hat{\alpha}_t} \epsilon_t, $ where $\hat{\alpha}$ represents constants hyperparameters, and  $\epsilon_t\sim \mathcal{N}(0, \mathbb{I})$ is noise at $t$ step.
\subsection{Boundary Enhance Conditioner and Core Enhance Conditioner}
\begin{figure}[t]
  \centering 
  \scalebox{0.43}{\includegraphics{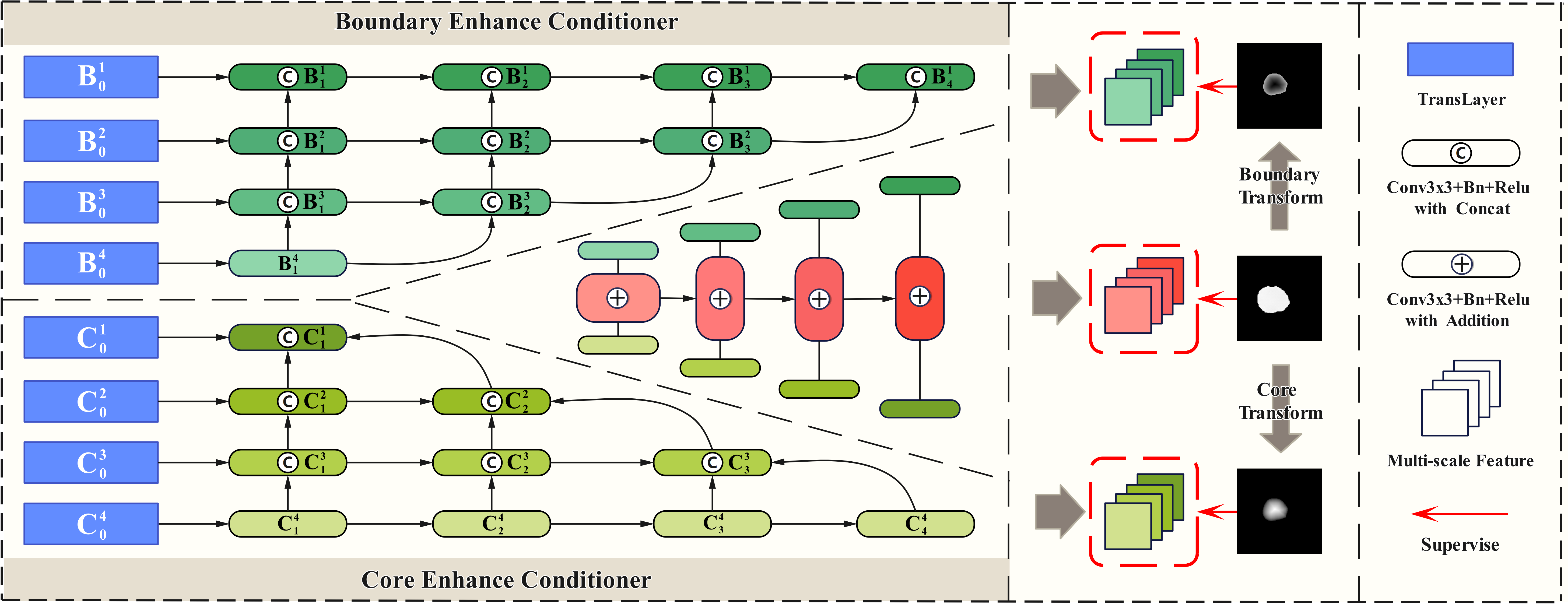}} % Adjust the scaling factor as needed
\caption{Detailed structures of our proposed BEC and CEC, which focus on learning the boundary and core information of the prostate under the guidance of decoupled soft labels.} \label{fig2}
\vspace{-0.2cm}
\end{figure}

Distinct from previous method that enhances edge learning via a weighted loss function, we proposed two parallel conditioners to decouple the learning of edge and core information.
As shown in Figure \ref{fig2}, the BEC starts with a higher number of convolutional layers then decreases as the network goes deep.
In contrast, the CEC increases the number of convolutional layers as the network deepens.
% The core concept drving this design is the use of an increased count of shallow convolutional layers within the BEC to reforce the learning of edge details. 
% While the CEC employs a dense stack of deeper convolutional layers to robustly capture the core information of the images.
Given that the sideouts of encoder are denoted as $f^{1}, f^{2}, f^{3}, f^{4}$ from large to small. 
Then these features at each level are transformed in parallel into a same number of dimensions (such as 64 in our implementation) via the $Trans$ layers. 
These layers follow by a combination of 3$\times$3 convolution, batchnorm and relu. 
Through these layers, we can obtain unified-channel features $B_{0}^{i}=Trans(f^{i})$ for $i$ from 1 to 4.
In this end, the multi-level features of the BEC at the $i$th row and $j$th colum $B_{j}^{i}$ can be denoted as:
\begin{equation}
  \begin{aligned}
    B_{j}^{i}=
  	\left\{
  	\begin{aligned}
   	&BConv(B_{j-1}^{i}\,\copyright\,Up(B_{j}^{i+1})), \quad \quad \quad \quad \quad if \; i+j\leq4, \\
    &BConv(B_{j-1}^{i}\,\copyright\,A), if \; i+j=5\, \& \,i=4, A=0; else \; A=Up(B_{j-1}^{i+1}), \\
  	\end{aligned}
  	\right.  
    % \\dsaW^{i}_{out}=Concat(W^{i}_{r \in \mathcal{R}}),
\end{aligned}
  \label{eq3}
  \end{equation}
where $BConv(\cdot)$ means 3 $\times$ 3 Conv-Bn-Relu operation.
$\copyright$ indicates the concatenation operation and $Up(\cdot)$ is an upsample operation with an upsampling rate 2.
Analogously, $C_{0}^{i}=Trans(f^{i}),\,i=1-4$. The implementation of the CEC at the $i$th row and $j$th colum $C_{j}^{i}$ formulated as:
\begin{equation}
  \begin{aligned}
    C_{j}^{i}=
  	\left\{
  	\begin{aligned}
    &BConv(C_{j-1}^{i}), \quad \quad \quad \quad \quad \quad \quad \quad \quad \quad \quad \quad \quad if \; i=4, \\
    &BConv(C_{j-1}^{i}\,\copyright\,Up(C_{j}^{i+1}) \,\copyright\,A), if \; i<4\, \& \, i=j, A=Up(C_{j+1}^{i+1});\\
    &\quad \quad \quad \quad \quad \quad \quad \quad \quad \quad \quad \quad \quad \quad \quad \quad \quad \quad \quad else \, A=0,\\
  	\end{aligned}
  	\right.  
    % \\dsaW^{i}_{out}=Concat(W^{i}_{r \in \mathcal{R}}),
\end{aligned}
  \label{eq4}
  \end{equation}
then the multi-scale features from the BEC and CEC are fed into a streamlined FPN to obtain the integrated prostate features $P^{i}$. It can be defined as:
\begin{equation}
  \begin{aligned}
    P^{i}=BConv(B_{5-i}^{i}\, \oplus C_{i}^{i}\, \oplus\, A),  if \;i=4, A=0; else \; A= Up(P^{i+1}),\\
    % \\dsaW^{i}_{out}=Concat(W^{i}_{r \in \mathcal{R}}),
\end{aligned}
  \label{eq5}
  \end{equation}
where $\oplus$ represents a pixel-wise summation operation. 
Finally, these multi-scale features will be supervised by a combing Dice Loss, BCE Loss, and IoU Loss, thus the total loss of our conditioners is:
\begin{equation}
  \mathcal{L}_{c} = \mathcal{L}_{bce}(B_{5-i}^{i}, g_{b})+\mathcal{L}_{bce}(C_{i}^{i}, g_{c})+\mathcal{L}_{bce}(P^{i}, g_{p}) +  \mathcal{L}_{IoU}(P^{i}, g_{p}) +  \mathcal{L}_{Dice}(P^{i}, g_{p}),
  % L_{\text{con}} = L_{\text{bce}}(B_{i}' \cdot \mathbb{1}_{\{i=j\}}, G_{B}) + L_{\text{bce}}(C_{i}' \cdot \mathbb{1}_{\{i=j\}}, G_{C}) + L_{\text{bce}}(P_{i}, G_{P}) + L_{\text{IoU}}(P_{i}, G)
  \label{eq6}
  \end{equation}
where $g_{b}$ and $g_{c}$ denote the boundary label and the core label of the prostate label $g_{p}$. 
We apply the Distance Transformation (DT)\cite{kimmel1996sub} on $g_{p}$ to differentiate between $g_{b}$ and $g_{c}$, obtaining a gradient image $I'$.
% The DT calculates the Euclidean distance from each foreground pixel to the nearest background pixel, obtaining a gradient image $I'$.
After normalization to [0,1] range,  pixels within the object's center exhibit the highest values, while those distant from the center or within the background display the lowest values.
Consequently, $I'$ reflects the central, more easily distinguishable aspects of the original image.
We then define the core label and the boundary label as $g_{c}=g_{p} \ast I' $, $g_{b}=g_{p} \ast (1-I')$, respectively.
\subsection{Crisscross Injection Strategy}
The proposed BEC and CEC are capable of capturing boundary and core features. 
However, directly injecting these features into the diffusion model in a stage-by-stage manner results in suboptimal feature utilization.
Thus, we proposed a crisscross injection strategy that allows the diffusion model to focus on object localization in early stages and refines the object's edges in later stages.
% Different from the generative tasks, the diffusion model for the segmentation incrementally adds noise to $g_{p}$ and learns the noise at time $t$ guided by image features.
% In the reverse stage, the diffusion model ultimately diffuses the final prediction from random Gaussian noise.
Following \cite{wu2023medsegdiff, wu2023medsegdiff2}, we adopted a modified ResUNet as our diffusion backbone and inject prostate features into the backbone layer by layer.
The encoder contains four convolutional stage with sequentially decreasing resolution.
Conversely, the decoder consists of four convolutional stages with sequentially increasing resolution.
We applied the cross-attention to facilitate the interaction between both boundary and core features with the diffusion feature.
Finally, the proposed Crisscross Injection Strategy can be denoted as:
\begin{equation}
  \begin{aligned}
    CIS=
  	\left\{
  	\begin{aligned}
    &E^{i}=CroAtt(A, E^{i}),\, if\, i=1,2, A=C_{i}^{i}; \, else \, A=B_{5-i}^{i}, \\
    &D^{5-i}=CroAtt(A, D^{5-i}),\, if\, i=1,2, A=C_{i}^{i}; \, else \, A=B_{5-i}^{i},\\
  	\end{aligned}
  	\right.  
    % \\dsaW^{i}_{out}=Concat(W^{i}_{r \in \mathcal{R}}),
  \end{aligned}
  \label{eq7}
\end{equation}
where $E^{i}$ means the $i$th outputs of encoder stage and $D^{i}$ is the $i$th outputs of decoder stage, respectively.
$CroAtt(\cdot)$ is the cross-attention operataion.
% The details of the Crisscross Injection Strategy can be found in the supplementary material.
%------------------------------table1-------------------	
\begin{table*}[t]
  \begin{center}
  %\vspace{-0.4cm
  % \vspace{0.001cm}
  \caption{Quantitative comparisons of DSC, IoU, HSD and ASD on three MRI datasets and one ultrasound datasets. For brevity, we denoted these metrics as D, I, H, and A, respectively. 
  The top two results are marked in $\textcolor{red}{red}$, \textcolor{blue}{blue}.}
  \label{table:1}
  \scalebox{0.75}{
  \begin{tabular}{c cccc cccc cccc cccc}
  \toprule[1pt]%\toprule
  \multirow{2}{*}[-3pt]{Method/Years}
  &\multicolumn{4}{c}{NCI-ISBI\cite{clark2013cancer}} 
  &\multicolumn{4}{c}{ProstateX\cite{litjens2014computer}}  
  &\multicolumn{4}{c}{Promise12\cite{litjens2014evaluation}} 
  &\multicolumn{4}{c}{CCH-TRUSPS\cite{feng2023multi}}
  % &\multicolumn{4}{c}{DUT-V2\cite{piao2020dutlfsaliency}}
    \\    
    \cmidrule(lr){2-5}  
    \cmidrule(lr){6-9} 
    \cmidrule(lr){10-13} 
    \cmidrule(lr){14-17}
  %   \cmidrule(lr){19-22}
    % \cmidrule(lr){23-26}
    &$D\uparrow$ &$I\uparrow$ &$H\downarrow$ &$A\downarrow$
    &$D\uparrow$ &$I\uparrow$ &$H\downarrow$ &$A\downarrow$
    &$D\uparrow$ &$I\uparrow$ &$H\downarrow$ &$A\downarrow$
    &$D\uparrow$ &$I\uparrow$ &$H\downarrow$ &$A\downarrow$
     % &$F_{\beta}\uparrow$ &$F_{\omega}\uparrow$ &$M\downarrow$&$E_{\gamma}\uparrow$
  \\ \midrule[0.5pt]
  % \hline
  Unet\cite{ronneberger2015u}$_{\rm{15MICCAI}}$ &.822	&.786	&2.32	&3.99	&.748	&.683	&3.41	&3.93	&.779	&.676	&4.25	&6.57	&.898	&.848	&5.58	&7.43 
  \\
  Unet++\cite{zhou2019unet++}$_{\rm{19TMI}}$ &.814	&.777	&2.30	&3.76	&.741	&.682	&3.44	&3.80	&.810	&.715	&4.12	&5.06	&.882	&.824	&5.92	&7.80 
  \\
  TransUnet\cite{chen2021transunet}$_{\rm{21Arxiv}}$ &.827	&.789	&2.28	&3.90	&.851	&.795	&\textcolor{blue}{2.92}	&2.33	&.887	&.812	&\textcolor{blue}{3.65}	&2.22	&.915	&.874	&\textcolor{blue}{5.36}	&\textcolor{blue}{4.43}  
  \\
  Swin-Unet\cite{cao2022swin}$_{\rm{22ECCV}}$ &.821	&.782	&2.39	&5.01	&.792	&.727	&3.32	&3.49	&.839	&.744	&4.09	&3.60	&.908	&.857	&5.78	&5.83  
  \\
  Uctransnet\cite{wang2022uctransnet}$_{\rm{22AAAI}}$ &.813	&.776	&2.38	&4.86	&.769	&.701	&3.40	&2.93	&.875	&.796	&3.87	&4.62	&.915	&.868	&5.61	&5.39 
  \\
  G-CASCADE\cite{rahman2024g}$_{\rm{24WACV}}$ &\textcolor{blue}{.842}	&\textcolor{blue}{.808}	&2.24	&3.75	&.844	&.795	&3.05	&2.02	&.880	&.802	&3.67	&2.67	&.915	&.871	&5.58	&5.93 
  \\ \midrule[0.5pt]
  CAT-Net\cite{hung2023cat}$_{\rm{23TMI}}$  & .841	&.810	&2.21	&4.04	&.796	&.743	&3.31	&2.67	&.888	&.813	&3.76	&2.51	&.895	&.850	&5.76	&5.01 
  \\
  CCT-Unet\cite{yan2023cct}$_{\rm{23JBHI}}$  &.836	&.803	&2.20	&4.50	&.803	&.756	&3.08	&2.22	&.857	&.775	&3.82	&3.51	&.902	&.852	&5.64	&6.98 
  \\
  MicroSegNet\cite{jiang2024microsegnet}$_{\rm{24CMIG}}$  &.829	&.796	&2.25	&3.86	&.849	&.798	&2.96	&2.45	&\textcolor{blue}{.890}	&\textcolor{blue}{.817}	&3.66	&\textcolor{blue}{2.19}	&\textcolor{red}{.928}	&\textcolor{red}{.886}	&5.49	&4.72 
  \\ \midrule[0.5pt]
  SegDiff\cite{amit2022segdiff}$_{\rm{21Arxiv}}$     & .807	&.776	&2.27	&4.12	&.835	&.788	&3.07	&\textcolor{blue}{1.93}	&		 -& -& -& -&.854	&.788	&5.86	&7.83   
  \\
  EnDiff\cite{wolleb2022ensemble}$_{\rm{22MIDL}}$     &.814	&.781	&2.32	&3.82	&.815	&.761	&3.24	&2.17   & -& -& -& - &.875	&.829	&6.04	&5.71  
  \\
  DermoSegDiff\cite{bozorgpour2023dermosegdiff}$_{\rm{23MICCAI}}$    & .841	&.806	&\textcolor{blue}{2.14}	&3.79	&\textcolor{blue}{.853}	&\textcolor{blue}{.804}	&2.96	&2.02	&  .885& .809& 3.69& 2.64&.900	&.855	&5.41	&4.59
  \\
  MedSegDiff-V2\cite{wu2023medsegdiff}$_{\rm{24AAAI}}$     &.828	&.796	&2.19	&\textcolor{blue}{3.71}	&.822	&.773	&3.10	&2.18	&.888	&.815	&3.67	&\textcolor{blue}{2.19}	&.844	&.772	&6.05	&8.55
  % \\ \midrule[0.5pt]                
  % Condition w/o Diffusion     & .831	&.791	&2.35	&5.27	&.835	&.769	&3.09	&2.56	&.855	&.773	&3.86	&3.01	&.895	&.856	&5.67	&5.09 
  \\
  Ours     & \textcolor{red}{.858}	&\textcolor{red}{.827}	&\textcolor{red}{2.04}	&\textcolor{red}{3.13}	&\textcolor{red}{.874}	&\textcolor{red}{.824}	&\textcolor{red}{2.86}	&\textcolor{red}{1.85}	&\textcolor{red}{.899}	&\textcolor{red}{.828}	&\textcolor{red}{3.63}	&\textcolor{red}{2.06}	&\textcolor{blue}{.923}	&\textcolor{blue}{.883}	&\textcolor{red}{5.35}	&\textcolor{red}{4.17}  
  % \\
  % \rowcolor{mygray}
  % $\Delta x$(\%)     & 3.26	&4.62	&13.1	&40.6	&4.65	&7.21	&7.25	&27.7	&5.14	&7.16	&5.89	&31.6	&3.15	&3.21	&5.69	&18.0
  \\
  \bottomrule[1pt]
  \vspace{-1cm}
  \end{tabular}}
  \end{center}
\end{table*}
\begin{figure}[t]
  \centering 
  \scalebox{0.47}{\includegraphics{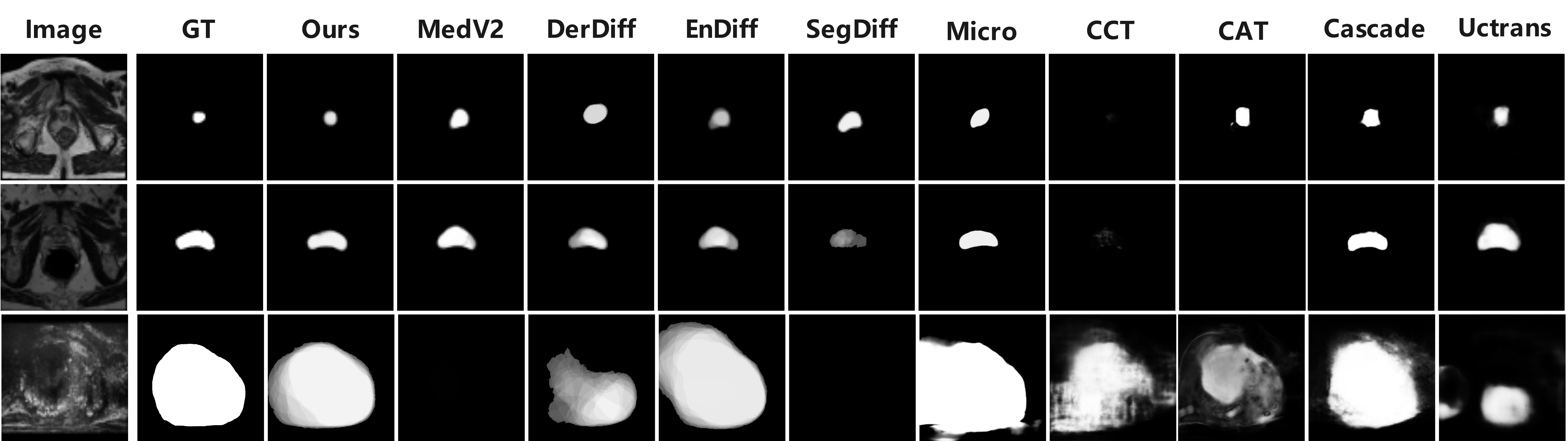}} % Adjust the scaling factor as needed
\caption{Visual comparisons of the proposed model and existing SOTA methods.} \label{fig3}
\end{figure}
\section{Experiments}
\subsection{Experiment Protocol}
\noindent
\textbf{Datasets.} We performed the evaluation on four public benchmark datasets, categorized into two types: three datasets comprising MRI images (NCI-ISBI\cite{clark2013cancer}, ProstateX\cite{litjens2014computer} and Promise12\cite{litjens2014evaluation}) and one dataset consisting of ultrasound images (CCH-TRUSPS\cite{feng2023multi}).
% NCI-ISBI\cite{clark2013cancer} is from ISBI 2013 Prostate Magnetic Resonance Imaging Challenge, containing 1571 T2WI images for training, 271 T2WI images for testing. 
% ProstateX\cite{litjens2014computer} includes 664 T2WI train images and 166 T2WI test images, which are performed by Siemens MAGNETOM Trio and Skyra 3T MR scanners.
% Promise12\cite{litjens2014evaluation} is from the Prostate MR Image Segmentation challenge, containing 778 T2WI train images and 418 T2WI test images.
% Prostate158\cite{adams2022prostate158} contains 3061 T2WI train images and 476 T2WI test images. Notably, the train images are annotated by experts and non-experts. The test images include clinical practitioners, graduate students, and both expert and non-expert annotators.
% For the fair comparison, we choose expert annotations for both train images and test images in this paper.
% CCH-TRUSPS\cite{feng2023multi} is an ultrasound prostate segmentation dataset collected from Chongqing University Cancer Hospital, containing 2152 ultrasound train images and 727 ultrasound test images.
Details of these datasets are provided in the supplementary material.
% All 3D scans are converted into 2D slices. Then, each slice is resized to 256 $\times$ 256 and normalized to $[0, 1]$ for training.
% To comprehensively validate the proposed model
\textbf{Metrics.} To validate the proposed model, we adopt four metrics: Dice Similarity Coefficient (DSC), Intersection over Union (IoU), Hausdorff Distance (HSD) and Average Surface Distance (ASD).
% \subsection{Implementation detail}
\textbf{Implementation detail.}
% We implemented the pre-train phase and training of network on the Pytorch toolbox with two RTX 4090 GPUs. 
% The encoder for feature extraction adopted in our method is PVT-B2\cite{wang2021pvtv2}. 
% The training procedure employs a batch size of 6 and utilizes the AdamW optimizer with a base learning rate of 1e-5. We run the model 25 times for the ensemble and T is 500. 
% For each dataset, we trained the network loading the architecture parameters $\theta$ with 100000 iterations, and we applied data augmentation techniques included horizontal and vertical random flipping and random cropping.
% The detail of the pre-train phase is shown in the supplementary material.
We trained our network using the Pytorch toolbox on two RTX 4090 GPUs, employing PVT-B2\cite{wang2021pvtv2} as the encoder. The training utilized a batch size of 6, the AdamW optimizer with a 1e-5 learning rate, and included 100,000 iterations. We conducted 25 ensemble runs with T=500. Pre-train phase details are in the supplementary material.
\begin{table*}[t]
  \centering
  \begin{minipage}{.55\linewidth}
    \centering
    \caption{Different conditioner settings.}
    \label{table:2}
    \scalebox{0.6}{
      \begin{tabular}{cccccccccccccccc}
        \toprule[1pt]%\toprule
        \multirow{2}{*}{Index} &
        \multicolumn{4}{c}{Conditioner} &
        \multirow{2}{*}{\begin{tabular}[c]{@{}c@{}}Param\\ Size(M)\end{tabular}} &
        \multicolumn{4}{c}{ProstateX\cite{litjens2014computer}} & 
        \multicolumn{4}{c}{CCH-TRUSPS\cite{feng2023multi}} \\
        \cmidrule(lr){2-5} \cmidrule(lr){7-10} \cmidrule(lr){11-14}
        & P & P$\ast$  & C &B & & $D\uparrow$ & $I\uparrow$ & $H\downarrow$ & $A\downarrow$ & $D\uparrow$ & $I\uparrow$ & $H\downarrow$ & $A\downarrow$ \\  
        \midrule
        (1)  & $\checkmark$  &  &  &  & 53.35  & .843  & .778  & 3.01  & 2.46  & .895  & .856  & 5.67  & 5.09
        \\
        (2) &  & $\checkmark$  &  &  & 53.36  & .857  & .799  & 2.95  & 2.21  & .906  & .867  & 5.61  & 4.85
        \\
        (3) & $\checkmark$  &  & $\checkmark$  &  & 54.06  & .852  & .801  & 2.95  & 2.26  & .909  & .871  & 5.56  & 4.45
        \\
        (4) & $\checkmark$  &  &  & $\checkmark$  & 54.13  & .865  & .811  & \textcolor{red}{2.85}  & 2.09  & .916  & .873 & 5.55 & 4.36
        \\
        (5) & $\checkmark$  &  & $\checkmark$  & $\checkmark$  & 54.63  & \textcolor{red}{.874}  & \textcolor{red}{.824}  & 2.86  & \textcolor{red}{1.85}  & \textcolor{red}{.923}  & \textcolor{red}{.883}  &\textcolor{red}{5.35}  & \textcolor{red}{4.17}
        \\
        \bottomrule
        \end{tabular}
    }
  \end{minipage}%
  \begin{minipage}{.45\linewidth}
    \centering
    \caption{Different init method comparison.}
    \label{table:4}
    \scalebox{0.7}{
      \begin{tabular}{ccccccccccc}
      \toprule
      \multirow{2}{*}{Method} & \multicolumn{4}{c}{ProstateX\cite{litjens2014computer}} & \multicolumn{4}{c}{CCH-TRUSPS\cite{feng2023multi}} \\
      \cmidrule(lr){2-5} \cmidrule(lr){6-9}
      & $D\uparrow$ & $I\uparrow$ & $H\downarrow$ & $A\downarrow$ & $D\uparrow$ & $I\uparrow$ & $H\downarrow$ & $A\downarrow$ \\
      \midrule
      Random & .865 & .812 & 2.92 & 1.91 & .893 & .858 & 5.68 & 4.72 \\
      Kaiming & .868 & .817 & 2.95 & 1.91 & .896 & .862 & 5.63 & 4.52 \\
      Ours & \textcolor{red}{.874} & \textcolor{red}{.824} & \textcolor{red}{2.86} & \textcolor{red}{1.85} & \textcolor{red}{.923} & \textcolor{red}{.883} & \textcolor{red}{5.35} & \textcolor{red}{4.17} \\
      \bottomrule
      \end{tabular}
    }
  \end{minipage}
\vspace{-0.2cm}
\end{table*}
% \vspace{-0.5mm}
% \vspace{l}
\subsection{Comparison with State-of-the-arts}
% The performance of the proposed method is compared with general medical image segmentation methods, prostate segmentation methods and  DPM-based Methods.
% Specifically,
% % , CCBANet\cite{nguyen2021ccbanet} and CFANet\cite{zhou2023cross} are proposed for Polyp Segmentation.
% % BA-Trans\cite{wang2021boundary} and DEU-Net\cite{karimi2023deu} are proposed for Skin Lesion Segmentation.
% % EDNnet\cite{wu2022edn} and SelfReformer\cite{SelfReformer} are proposed for Salient Object Detection.
% CAT-Net\cite{hung2023cat}, CCT-Unet\cite{yan2023cct} and MicroSegNet\cite{jiang2024microsegnet} are proposed for Prostate Segmentation.
% Unet\cite{ronneberger2015u}, Unet++\cite{zhou2019unet++}, TransUnet\cite{chen2021transunet}, Swin-Unet\cite{cao2022swin}, Uctransnet\cite{wang2022uctransnet} and G-CASCADE\cite{rahman2024g} are proposed for General Medical Segmentation. 
% We also compared 4 DPM-based methods included SegDiff\cite{wolleb2022ensemble}, EnDiff\cite{wolleb2022ensemble}, DermoSegDiff\cite{bozorgpour2023dermosegdiff} and MedSegDiff-V2\cite{wu2023medsegdiff}. 
% For a fair comparison, we replaced the encoder of these methods with PVT-B2\cite{wang2021pvtv2}. We train and infer these methods for the same number of iterations and ensemble times as our model.
The performance of the proposed method is compared with 6 general medical image segmentation methods, including Unet\cite{ronneberger2015u}, Unet++\cite{zhou2019unet++}, TransUnet\cite{chen2021transunet}, Swin-Unet\cite{cao2022swin}, Uctransnet\cite{wang2022uctransnet} and G-CASCADE\cite{rahman2024g}.
We also compared 3 prostate segmentation methods, including CAT-Net\cite{hung2023cat}, CCT-Unet\cite{yan2023cct} and MicroSegNet\cite{jiang2024microsegnet}.
Additionally, we compared 4 DPM-based methods included SegDiff\cite{amit2022segdiff}, EnDiff\cite{wolleb2022ensemble}, DermoSegDiff\cite{bozorgpour2023dermosegdiff} and MedSegDiff-V2\cite{wu2023medsegdiff}. 
% including Unet\cite{ronneberger2015u}, Unet++\cite{zhou2019unet++}, TransUnet\cite{chen2021transunet}, Swin-Unet\cite{cao2022swin}, Uctransnet\cite{wang2022uctransnet}, G-CASCADE\cite{rahman2024g},
% CAT-Net\cite{hung2023cat}, CCT-Unet\cite{yan2023cct} and MicroSegNet\cite{jiang2024microsegnet}.
% We also compared 4 DPM-based methods included SegDiff\cite{wolleb2022ensemble}, EnDiff\cite{wolleb2022ensemble}, DermoSegDiff\cite{bozorgpour2023dermosegdiff} and MedSegDiff-V2\cite{wu2023medsegdiff}. 
% , CCBANet\cite{nguyen2021ccbanet} and CFANet\cite{zhou2023cross} are proposed for Polyp Segmentation.
% BA-Trans\cite{wang2021boundary} and DEU-Net\cite{karimi2023deu} are proposed for Skin Lesion Segmentation.
% EDNnet\cite{wu2022edn} and SelfReformer\cite{SelfReformer} are proposed for Salient Object Detection.
% CAT-Net\cite{hung2023cat}, CCT-Unet\cite{yan2023cct} and MicroSegNet\cite{jiang2024microsegnet}
% Unet\cite{ronneberger2015u}, Unet++\cite{zhou2019unet++}, TransUnet\cite{chen2021transunet}, Swin-Unet\cite{cao2022swin}, Uctransnet\cite{wang2022uctransnet} and G-CASCADE\cite{rahman2024g} are proposed for General Medical Segmentation. 
% We also compared 4 DPM-based methods included SegDiff\cite{wolleb2022ensemble}, EnDiff\cite{wolleb2022ensemble}, DermoSegDiff\cite{bozorgpour2023dermosegdiff} and MedSegDiff-V2\cite{wu2023medsegdiff}. 
For a fair comparison, we replaced the encoder of these DPM-based methods with PVT-B2\cite{wang2021pvtv2}, except for the EnDiff, which does not utilize the encoder. 
We trained and inferred these methods for the same number of iterations and ensemble times as our model.
\textbf{Results.}  
% Table \ref{table:1} shows the quantitative comparison results between our model and previous SOTA methods.
% As shown in Table \ref{table:1}, the condition means the performance of the three conditions without the diffusion model. 
% As shown in Table \ref{table:1}, our method outperform other methods across four datasets, except for second-best results in Dice and IoU metrics on the CCH-TRUSPS, with only minor differences at the thousandth digit compared to the best method.
As shown in Table \ref{table:1}, our method improves the IoU by an average of 2.1\% and reduces the ASD by an average of 8.5\% across three MRI prostate datasets compared to the second-best method. On the ultrasound prostate dataset, the Dice and IoU metrics show only minor differences at the thousandth digit compared to the best method. This may be due to the simplicity of this dataset, where multiple ensemble runs might introduce noise, slightly reducing performance.
% our methods inject the boundary and core features provided by the conditions, achieving substantial improvements, particularly under the NCI-ISBI and Promise12 datasets, with the ASD metric increasing by 40.6\% and 31.6\%, respectively, while maintaining superior performance compared to other methods.
Intuitively, We visualize segmentation maps from our model and others in Figure \ref{fig3}.
It is obvious that our model not only achieves precise localization but also clearly delineates boundaries for prostates of varying sizes. 
% produces boundary of various prostate sizes  and precise localization on prostates of various sizes, including cluttered backgrounds (Row 4-6)
% showing our method's effective performance on prostates of various sizes.
% Our method excels in dealing with various challenging scenarios, including cluttered backgrounds (Row 4-6), Multiple objects (Row 7).
More visualized results can be found in the supplementary material.
\begin{figure}[t]
  \centering
  \begin{minipage}{.5\linewidth}
    \captionof{table}{Injection strategy comparison.}
    \label{table:3}
    \scalebox{0.65}{
      \begin{tabular}{ccccccccccccccc}
      \toprule
      \multirow{2}{*}{Index} & \multicolumn{4}{c}{Strategy} & \multicolumn{4}{c}{ProstateX\cite{litjens2014computer}} & \multicolumn{4}{c}{CCH-TRUSPS\cite{feng2023multi}} \\
      \cmidrule(lr){2-5} \cmidrule(lr){6-9} \cmidrule(lr){10-13}
      & SbS & 2:2 & 1:3 & 3:1 & $D\uparrow$ & $I\uparrow$ & $H\downarrow$ & $A\downarrow$ & $D\uparrow$ & $I\uparrow$ & $H\downarrow$ & $A\downarrow$ \\
      \midrule
      (1) & $\checkmark$ & & & & .856 & .806 & 2.95 & 1.96 & .903 & .862 & 5.53 & 4.73 \\
      (2) & & $\checkmark$ & & & .874 & .824 & \textcolor{red}{2.86} & \textcolor{red}{1.85} & .923 & .883 & \textcolor{red}{5.35} & \textcolor{red}{4.17} \\
      (3) & & & $\checkmark$ & & \textcolor{red}{.875} & .820 & 2.89& 1.99 & \textcolor{red}{.927} & .875 &5.45 &4.45\\
      (4) & & & & $\checkmark$ & .865 & .816 & 2.91& 1.95&  .923 &\textcolor{red}{.888} &5.32 &4.26 \\
      \bottomrule
      \end{tabular}
    }
  \end{minipage}%
  \begin{minipage}{.5\linewidth}
    \centering
    \scalebox{0.9}{
    \includegraphics[width=\linewidth]{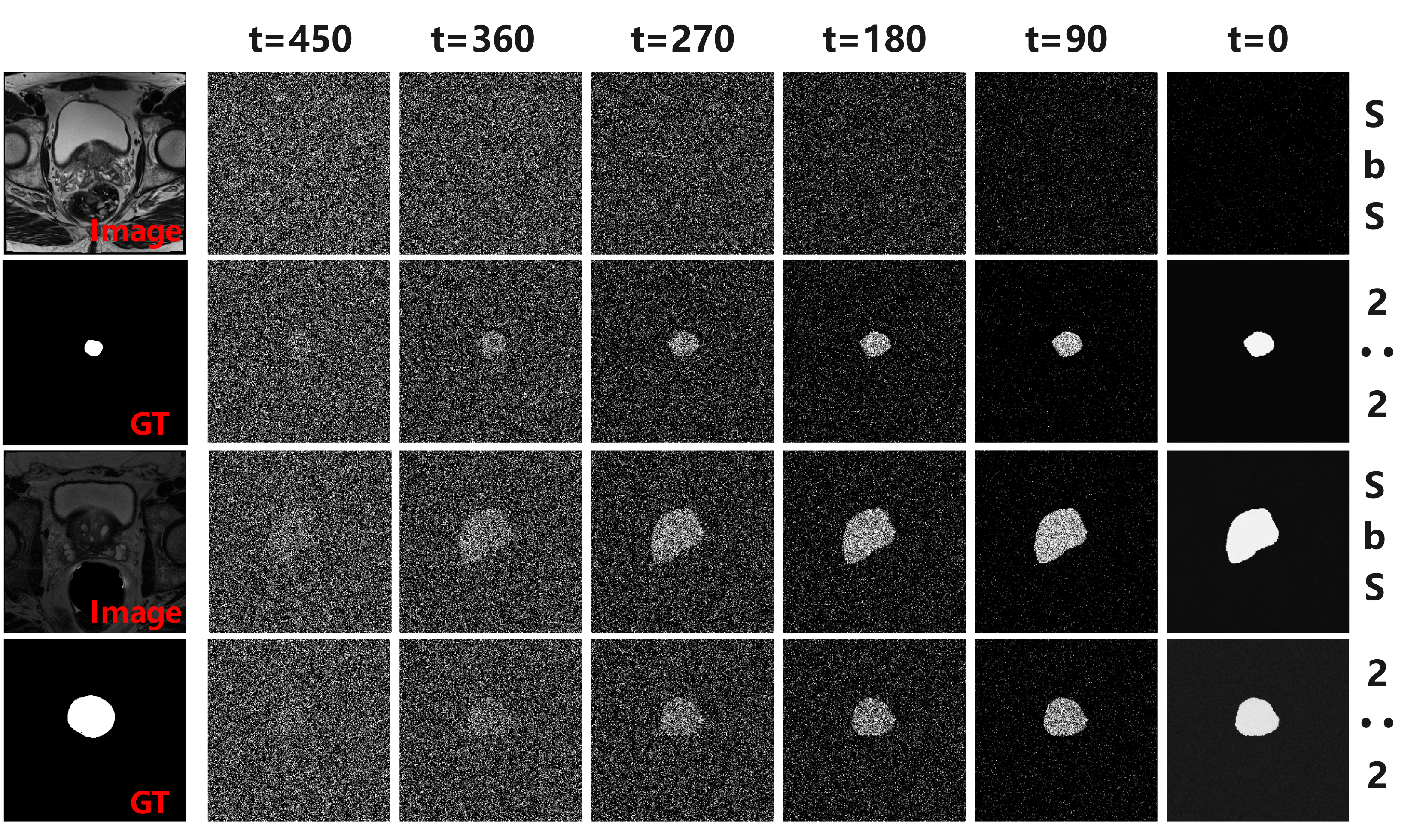}} % replace with your image name
    \captionof{figure}{Visual comparisons with (1) and (2) are shown in Table \ref{table:3}.}
    \label{fig:image1}
  \end{minipage}
  \label{fig14}
  \vspace{-0.5cm}
\end{figure}
% \vspace{-0.3cm}
\subsection{Ablation Study}
\textbf{Effect of BEC and CEC.}
We validated our proposed conditioner structures through five ablation studies.
% We conducted five ablation experiments to validate the effectiveness of our proposed conditioner structures. Specifically, we manually adjusted the setting of different conditioner in Table \ref{table:2}.
As shown in Table \ref{table:2}, 
% \vspace{-5mm}
% \vspace{-0.1mm}
(1) only prostate conditioner to inject.
(2) a simple FPN replaced three conditioners, predicting three-channel features representing prostate, core and boundary features for injecting.
(3) prostate and core conditioners to inject.
(4) prostate and boundary conditioners to inject.
(5) all three conditioners to inject.
When injecting the same features, we observed that our proposed boundary and core conditioners significantly enhanced model performance compared with (2) and (5). 
These results demonstrate decoupled learning of boundary and core features can more effectively improve performance with a smaller model size.
Moreover, compared with (1), (3), (4) and (5), the absence of either one or both boundary and core conditioners for learning and injecting features resulted in a decrease in Dice scores on the ProstateX and CCH-TRUSPS datasets, respectively. 
This further highlights the effectiveness of the boundary and core conditioner architectures.
% From Table \ref{table:2} (1), (3), (4) and (5), we observed that our proposed injection structure achieved significant performance improvements compared to traditional injection structures, whether injecting body or detail individually or injecting both.
% Moreover, to validate the effectiveness of our designed injection structure, we compared it with a simple application of the same concept, as shown in Table \ref*{table:2} (2) and (5).
% It is worth noting that our proposed method achieved performance improvements of 1.7\% and 1.8\% in Dice scores on the ProstateX\cite{litjens2014computer} and MUPS\cite{feng2023multi} datasets, respectively, with an increase of only 1.27M parameters.
% These results powerful indicate that our proposed injection structure achieved outstanding performance improvements with the addition of a small number of parameters, thereby demonstrating that efficiently utilizing high and low-level information can lead to effective performance enhancements.
\\
\textbf{Effect of CIS.}
We performed four quantitative experiments to validate the effectiveness of CIS. 
% Specifically, we manually adjusted the feature injection stage for different setting in Table \ref{table:3}.
As shown in Table \ref{table:3}, 2:2 means a ratio of injection layer for using the proposed strategy to inject core features into the shallow two layers and detail features into the deeper two layers of both encoder and decoder.
% (2) adopts the proposed strategy, maintaining a 2:2 ratio of layer for injecting core features into the shallow two layers and detail features into the deeper two layers of both encoder and decoder.
% In contrast, (1) stage by stage injects (SbS) both boundary features and core features into shallow layers and deeper layer.
% (3) and (4) changes the ratio between shallow and deeper layers is to 1:3 and 3:1.
SbS denotes that the stage by stage strategy injects boundary features and core features into shallow layers and deeper layers with a 2:2 ratio.
Compared (1) with (2), (3), and (4), we observed that our proposed injection strategy significantly outperforms the traditional stage-by-stage injection approach, regardless of whether using a 3:1 or 1:3 ratio. 
These results strongly validate the adequacy of utilizing multi-level features in our injection strategy.
To further illustrate the qualitative effect of our strategy, we visualized feature maps of (1) and (2) in Figure \ref{fig:image1}. 
It can be seen that the proposed strategy enables the diffusion model to focus on object localization, preventing entirely black predictions, especially when the prostate region is small. 
Simultaneously, it allows the model to focus on edge areas achieving precise edge segmentation.
% From Table \ref*{table:3} (1) and (2), it can be observed that our proposed criss-cross injection strategy outperforms the conventional injection strategy in terms of performance.
% Furthermore, to further validate the adequacy of the information utilization from the proposed injection strategy, we adopted different ratios between high and low-level divisions. As shown in Table \ref*{table:3} (2), (3) and (4), it can be observed that our proposed injection strategy leads to performance improvements.
% Obviously, these results confirm the superiority of the proposed criss-cross injection strategy. 
\\
\textbf{Effect of GP.}
To validate the effectiveness of GP, we conducted a set of experiments over different initialization methods (Random and Kaiming).
As shown in Table \ref{table:4}, our method demonstrates superior performance across two datasets on four metrics, especially showing 3.3\% and 2.8\% improvements in Dice and IoU on the CCH-TRUSPS dataset, thereby affirming the benefits of generative pre-training for model initialization.
Through the proposed GP, the diffusion model acquire the capability to represent prostate features, bridging the domain gap between the conditional features and the diffusion model features. 
To illustrate this point, we showcased some generated prostate images in the supplementary material.
These images clearly possess structures characteristic of prostate imagery and closely resemble real prostate images.
\\
\section{Conclusion}
In this paper, we proposed CriDiff,  a novel framework for prostate segmentation, which efficiently learns and injects multi-scale edge and non-edge features into the diffusion network using two parallel conditioners (BEC and CEC) and a crisscross injection strategy (CIS).
% In this paper, we proposed a novel framework (CriDiff) for prostate segmentation.
% It efficiently learns edge and non-edge information of images and effectively injects their multi-scale features into the diffusion network by leveraging proposed two parallel conditioners and a crisscross injection strategy.
To bridge the domain gap between image and diffusion model features, we introduced a generative method without introducing additional parameters.
Experimental results demonstrate that our proposed method can achieve state-of-the-art performance in prostate segmentation.
% The Boundary Enhance Conditioner (BEC) and Core Enhance Conditioner (CEC) decouples image features and learns complementary edge and non-edge features. 
\begin{credits}
\subsubsection{\ackname} This work was supported by the National Natural Science Foundation of China (62376050, 62372080, 62172070, and U22B2052), and the Dalian Science and Technology Innovation Foundation (2023JJ11CG001 and 2022JJ11CG001).

\subsubsection{\discintname}
We declared no competing interests.
\end{credits}
\bibliographystyle{splncs04}
\bibliography{Paper-0339}

\begin{thebibliography}{10}
\providecommand{\url}[1]{\texttt{#1}}
\providecommand{\urlprefix}{URL }
\providecommand{\doi}[1]{https://doi.org/#1}

\bibitem{amit2022segdiff}
Amit, T., Shaharbany, T., Nachmani, E., Wolf, L.: Segdiff: Image segmentation with diffusion probabilistic models. arXiv preprint arXiv:2112.00390  (2021)

\bibitem{bozorgpour2023dermosegdiff}
Bozorgpour, A., Sadegheih, Y., Kazerouni, A., Azad, R., Merhof, D.: Dermosegdiff: A boundary-aware segmentation diffusion model for skin lesion delineation. In: International Workshop on PRedictive Intelligence In MEdicine. pp. 146--158. Springer (2023)

\bibitem{cao2022swin}
Cao, H., Wang, Y., Chen, J., Jiang, D., Zhang, X., Tian, Q., Wang, M.: Swin-unet: Unet-like pure transformer for medical image segmentation. In: European conference on computer vision. pp. 205--218. Springer (2022)

\bibitem{chen2021transunet}
Chen, J., Lu, Y., Yu, Q., Luo, X., Adeli, E., Wang, Y., Lu, L., Yuille, A.L., Zhou, Y.: Transunet: Transformers make strong encoders for medical image segmentation. arXiv preprint arXiv:2102.04306  (2021)

\bibitem{clark2013cancer}
Clark, K., Vendt, B., Smith, K., Freymann, J., Kirby, J., Koppel, P., Moore, S., Phillips, S., Maffitt, D., Pringle, M., et~al.: The cancer imaging archive (tcia): maintaining and operating a public information repository. Journal of digital imaging  \textbf{26},  1045--1057 (2013)

\bibitem{feng2023multi}
Feng, Y., Atabansi, C.C., Nie, J., Liu, H., Zhou, H., Zhao, H., Hong, R., Li, F., Zhou, X.: Multi-stage fully convolutional network for precise prostate segmentation in ultrasound images. Biocybernetics and Biomedical Engineering  \textbf{43}(3),  586--602 (2023)

\bibitem{guo2015deformable}
Guo, Y., Gao, Y., Shen, D.: Deformable mr prostate segmentation via deep feature learning and sparse patch matching. IEEE transactions on medical imaging  \textbf{35}(4),  1077--1089 (2015)

\bibitem{ho2020denoising}
Ho, J., Jain, A., Abbeel, P.: Denoising diffusion probabilistic models. Advances in neural information processing systems  \textbf{33},  6840--6851 (2020)

\bibitem{hung2023cat}
Hung, A.L.Y., Zheng, H., Miao, Q., Raman, S.S., Terzopoulos, D., Sung, K.: Cat-net: A cross-slice attention transformer model for prostate zonal segmentation in mri. IEEE transactions on medical imaging  \textbf{42}(1),  291--303 (2023)

\bibitem{jiang2024microsegnet}
Jiang, H., Imran, M., Muralidharan, P., Patel, A., Pensa, J., Liang, M., Benidir, T., Grajo, J.R., Joseph, J.P., Terry, R., et~al.: Microsegnet: A deep learning approach for prostate segmentation on micro-ultrasound images. Computerized Medical Imaging and Graphics p. 102326 (2024)

\bibitem{kimmel1996sub}
Kimmel, R., Kiryati, N., Bruckstein, A.M.: Sub-pixel distance maps and weighted distance transforms. Journal of Mathematical Imaging and Vision  \textbf{6},  223--233 (1996)

\bibitem{litjens2014computer}
Litjens, G., Debats, O., Barentsz, J., Karssemeijer, N., Huisman, H.: Computer-aided detection of prostate cancer in mri. IEEE transactions on medical imaging  \textbf{33}(5),  1083--1092 (2014)

\bibitem{litjens2014evaluation}
Litjens, G., Toth, R., Van De~Ven, W., Hoeks, C., Kerkstra, S., Van~Ginneken, B., Vincent, G., Guillard, G., Birbeck, N., Zhang, J., et~al.: Evaluation of prostate segmentation algorithms for mri: the promise12 challenge. Medical image analysis  \textbf{18}(2),  359--373 (2014)

\bibitem{pellicer2022deep}
Pellicer-Valero, O.J., Marenco~Jimenez, J.L., Gonzalez-Perez, V., Casanova Ramon-Borja, J.L., Martin~Garcia, I., Barrios~Benito, M., Pelechano~Gomez, P., Rubio-Briones, J., Rup{\'e}rez, M.J., Mart{\'\i}n-Guerrero, J.D.: Deep learning for fully automatic detection, segmentation, and gleason grade estimation of prostate cancer in multiparametric magnetic resonance images. Scientific reports  \textbf{12}(1), ~2975 (2022)

\bibitem{rahman2024g}
Rahman, M.M., Marculescu, R.: G-cascade: Efficient cascaded graph convolutional decoding for 2d medical image segmentation. In: Proceedings of the IEEE/CVF Winter Conference on Applications of Computer Vision. pp. 7728--7737 (2024)

\bibitem{ronneberger2015u}
Ronneberger, O., Fischer, P., Brox, T.: U-net: Convolutional networks for biomedical image segmentation. In: Medical Image Computing and Computer-Assisted Intervention--MICCAI 2015: 18th International Conference, Munich, Germany, October 5-9, 2015, Proceedings, Part III 18. pp. 234--241. Springer (2015)

\bibitem{siegel2020cancer}
Siegel, R.: Cancer statistics, 2020. CA: a cancer journal for clinicians.  \textbf{70}(1), ~7 (2020)

\bibitem{tian2015superpixel}
Tian, Z., Liu, L., Zhang, Z., Fei, B.: Superpixel-based segmentation for 3d prostate mr images. IEEE transactions on medical imaging  \textbf{35}(3),  791--801 (2015)

\bibitem{wang2022uctransnet}
Wang, H., Cao, P., Wang, J., Zaiane, O.R.: Uctransnet: rethinking the skip connections in u-net from a channel-wise perspective with transformer. In: Proceedings of the AAAI conference on artificial intelligence. vol.~36, pp. 2441--2449 (2022)

\bibitem{wang2021pvtv2}
Wang, W., Xie, E., Li, X., Fan, D.P., Song, K., Liang, D., Lu, T., Luo, P., Shao, L.: Pvtv2: Improved baselines with pyramid vision transformer. Computational Visual Media  \textbf{8}(3),  1--10 (2022)

\bibitem{warfield2002validation}
Warfield, S.K., Zou, K.H., Wells, W.M.: Validation of image segmentation and expert quality with an expectation-maximization algorithm. In: Medical Image Computing and Computer-Assisted Intervention—MICCAI 2002: 5th International Conference Tokyo, Japan, September 25--28, 2002 Proceedings, Part I 5. pp. 298--306. Springer (2002)

\bibitem{wolleb2022ensemble}
Wolleb, J., Sandk{\"u}hler, R., Bieder, F., Valmaggia, P., Cattin, P.C.: Diffusion models for implicit image segmentation ensembles. In: International Conference on Medical Imaging with Deep Learning. pp. 1336--1348. PMLR (2022)

\bibitem{wu2023medsegdiff2}
Wu, J., Fu, R., Fang, H., Zhang, Y., Xu, Y.: Medsegdiff-v2: Diffusion based medical image segmentation with transformer. arXiv preprint arXiv:2301.11798  (2023)

\bibitem{wu2023medsegdiff}
Wu, J., FU, R., Fang, H., Zhang, Y., Yang, Y., Xiong, H., Liu, H., Xu, Y.: Medsegdiff: Medical image segmentation with diffusion probabilistic model. In: Medical Imaging with Deep Learning (2023)

\bibitem{yan2023cct}
Yan, Y., Liu, R., Chen, H., Zhang, L., Zhang, Q.: Cct-unet: A u-shaped network based on convolution coupled transformer for segmentation of peripheral and transition zones in prostate mri. IEEE Journal of Biomedical and Health Informatics  (2023)

\bibitem{yu2017volumetric}
Yu, L., Yang, X., Chen, H., Qin, J., Heng, P.A.: Volumetric convnets with mixed residual connections for automated prostate segmentation from 3d mr images. In: Proceedings of the AAAI Conference on Artificial Intelligence. vol.~31 (2017)

\bibitem{zhou2019unet++}
Zhou, Z., Siddiquee, M.M.R., Tajbakhsh, N., Liang, J.: Unet++: Redesigning skip connections to exploit multiscale features in image segmentation. IEEE transactions on medical imaging  \textbf{39}(6),  1856--1867 (2019)

\end{thebibliography}
\end{document}